\documentclass{sig-alternate}
\pdfoutput=1

\usepackage{graphicx}
\usepackage{array}
\usepackage{balance}
\usepackage{color}
\usepackage{url}
\usepackage{hyperref}
\usepackage{topcapt}
\usepackage{float}
\usepackage{xspace}
\usepackage{booktabs}
\usepackage{multirow}

\usepackage{amssymb}
\usepackage{amsmath}

\usepackage{pgf}
\usepackage{tikz}
\usetikzlibrary{matrix}

\usepackage{url}

\usetikzlibrary{calc,trees,positioning,arrows,chains,shapes.geometric,%
    decorations.pathreplacing,decorations.pathmorphing,shapes,%
    matrix,shapes.symbols,topaths}

\begin{document}


\title{Benchmarking Soundtrack Recommendation\\ Systems with SRBench\titlenote{This work has been supported by the Excellence Cluster on Multimodal Computing and Interaction (MMCI).}}

\numberofauthors{2}
\author{
\alignauthor
Aleksandar Stupar\\
       \affaddr{Saarland University}\\
       \affaddr{Saarbr{\"u}cken, Germany}\\
       \email{astupar@mmci.uni-saarland.de}
\alignauthor
Sebastian Michel\\
       \affaddr{Saarland University}\\
       \affaddr{Saarbr{\"u}cken, Germany}\\
       \email{smichel@mmci.uni-saarland.de}
}

\maketitle
\begin{abstract}
In this work, a benchmark to evaluate the retrieval performance  of soundtrack recommendation systems is proposed. Such systems aim at finding  songs
that are played as background music for a given set of images. The proposed benchmark is based on preference judgments, where
relevance is considered a continuous ordinal variable and  judgments are collected for pairs of songs with respect to a query (i.e., set of images).
To capture a wide variety of songs and images, we  use a large space of possible music genres,
different emotions expressed through music, and various query-image themes.
The benchmark consists of two types of relevance assessments: (i) judgments obtained from a
user study, that serve as a ``gold standard'' for (ii)  relevance judgments gathered through
Amazon's Mechanical Turk. We report on an analysis of relevance judgments based on different levels of user agreement 
and investigate the performance of two state-of-the-art soundtrack recommendation systems
using the proposed benchmark.
\end{abstract}

\section{Introduction}
\label{sec:introduction}

With the increase of available multimedia content, searching for information contained in images, 
speech, music, or videos became an integral part in the information retrieval field.
Evaluating the quality of such systems introduces
new challenges as interpreting abstract associations---such as similarity between images---is complex
and can be done in various ways. 
Similar to the text retrieval evaluations considered in TREC~\cite{DBLP:journals/cacm/Voorhees07,link:trec}, 
evaluating music, image, and video retrieval systems has been the main concern of venues such as 
MIREX~\cite{link:mirex}, TRECVID \cite{link:trecvid}, and ImageCLEF \cite{link:imageclef}.

One of the largest contributions made by these venues is the proposal and standardization of a
retrieval corpus, i.e., the standardization of a document and a query collection. In addition to this, the defined benchmarks contain
human relevance judgments that assess the quality of query results with respect to the information need as described by the query.

In order to estimate standard retrieval measures, such as precision and recall,
it is essential for these assessments to be complete and reusable. If constructed properly, they enable a
fair and unbiased comparison among systems---which in turn increases competition and the pace of improvements in the field.
\vspace{+2mm}

\begin{figure}[t]
\centering
\includegraphics[width=0.95\columnwidth]{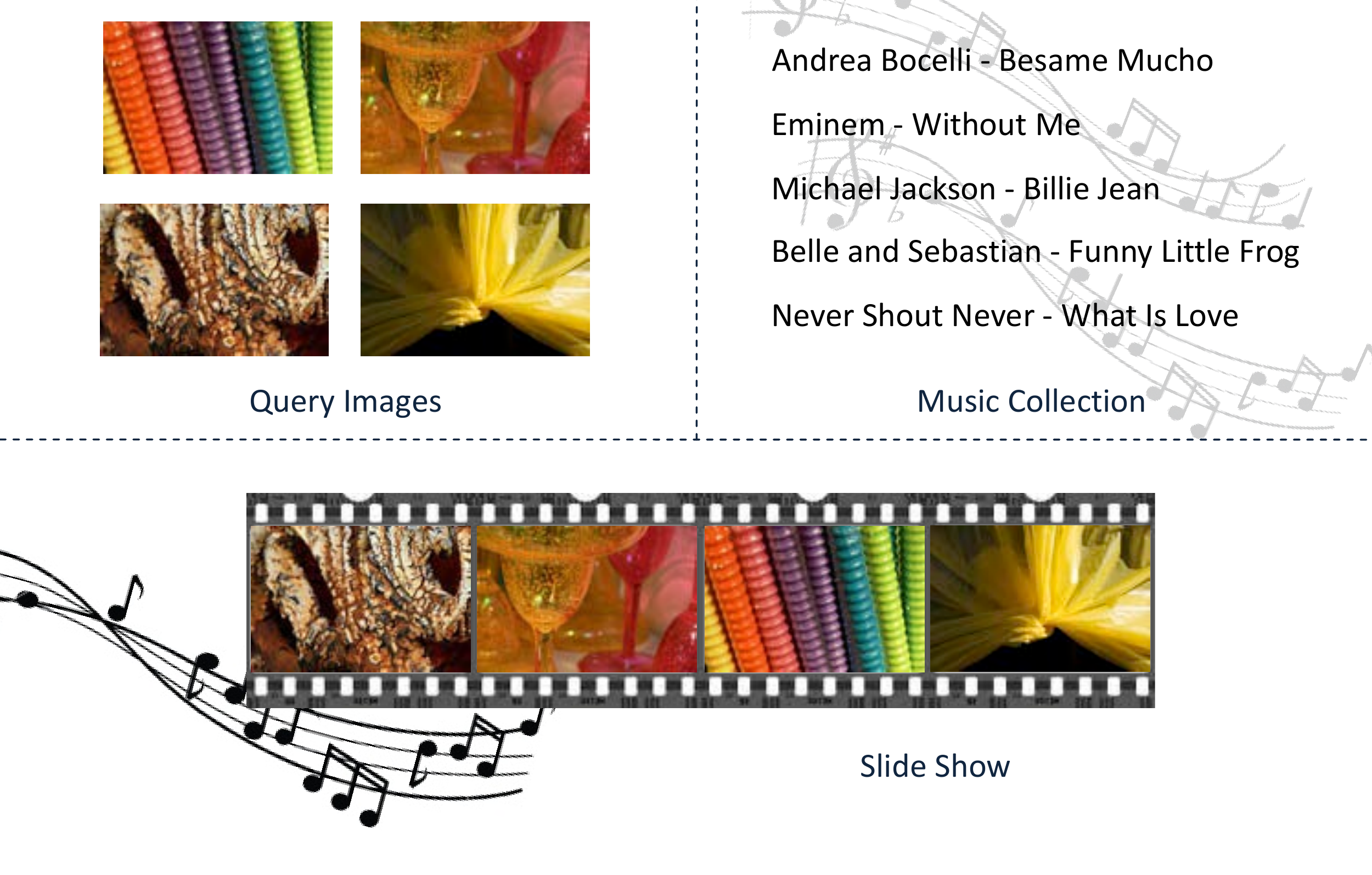}
\caption{Recommending music for slideshow images}
\label{fig:recommendation}
\end{figure}

In this work, we consider the problem of creating a benchmark dataset that is
used to assess the quality of soundtrack recommendation systems.
A soundtrack recommendation system is an information retrieval system that 
recommends a list of songs for a given set of images. The goal of the system is to 
retrieve the best matching songs to be used as background music during the presentation 
of images, as illustrated in Figure~\ref{fig:recommendation}.
The main difference to traditional information retrieval tasks is that, here,
queries and documents originate from completely different domains; queries are given in form of images
while the ``documents'' are music pieces.

As of now, the soundtrack recommendation state-of-the-art consists of only two approaches, namely an approach
by Li and Shan~\cite{DBLP:conf/mm/LiS07} and our own Picasso approach~\cite{DBLP:conf/sigir/StuparM11}.
We investigate and report on the performance of these two approaches using the proposed benchmark.
However, the benefit of a re-usable benchmark is far greater as it can improve the efficiency of continuous 
re-evaluations and comparisons of the existing approaches in the future. We expect that an open benchmark 
will, thus, foster the research on soundtrack recommendation systems.

\subsection{Problem Formulation and Outline}
\label{sec:problem}
A soundtrack recommendation system, over a set of indexed songs 
$S=\{s_{1}, s_{2},...\}$, takes as input/query a set of images $q=\{img_{1}, img_{2},...\}$ 
and the size of the result set $K$. It returns a subset of the indexed songs $S_{r} \subseteq  S$, with  $|S_{r}|=K$,  ordered 
with respect to their relevance to act as background music for a slideshow that features the given query images.

In this work, we address the problem of creating a benchmark to evaluate the retrieval performance of a given
soundtrack recommendation system. The proposed benchmark $B=(Q, S, R)$ contains a set of queries $Q=\{q_{1}, q_{2},...\}$,
a set of songs $S=\{s_{1}, s_{2},...\}$,  and a set of human relevance judgments $R=\{r_{1}, r_{2},...\}$, with each query $q_{i}$ defined as a set 
of images $q_{i}=\{img_{1}, img_{2},...\}$. 
The proposed benchmark fulfills the following important requirements:
\begin{itemize}
\item{it enables an unbiased comparison between different recommendation systems}
\item{it is reusable, that is, once created it can be used to evaluate systems with no additional human intervention}
\item{it provides high coverage in terms of ``document'' collection (songs) and evaluated queries (images) }
\item{it contains judgments with high agreement between assessors} 
\item{it is publicly available\footnote{\url{https://sites.google.com/site/srbench/}}}
\end{itemize}

Intuitively, the task of soundtrack recommendation appears to be highly subjective as the taste in music largely varies. However, 
as we will see, the agreement level between the assessors is quite high, indicating that it makes sense to address the problem
for the general case, i.e., to recommend soundtracks for the ``average'' user. 
It is important to note that the proposed benchmark can also be used to evaluate personalized recommendation systems 
where the evaluation is performed with respect to assessments of the individual assessors.

The rest of the paper is organized as follows: Section~\ref{sec:relatedwork} discusses related work. 
Section~\ref{sec:dataset} describes the used document collection and the queries. Section~\ref{sec:relevance} shows
how relevance assessments are used to measure retrieval effectiveness. Section~\ref{sec:building} explains the 
process of collecting relevance assessments and elaborates on various statistics of the collected assessments.
Section~\ref{sec:evaluation} reports on the results of the evaluation concerning the state-of-the-art approaches 
using the proposed benchmark. Section~\ref{sec:conclusion} concludes the paper. 

\section{Related Work}
\label{sec:relatedwork}

Our approach is based on the notion of pairwise comparisons, first mentioned and analyzed by 
Fechner~\cite{fecherelemente} and made popular later by Thurstone~\cite{thurstonelaw}.
Thurstone~\cite{thurstonelaw} used them to determine
the scale of perceived stimuli and referred to it as the {\it law of comparative judgment}.
A large body of research exists on reconstructing the final ranking from a set of pairwise 
comparisons~\cite{DBLP:conf/sigir/CarteretteP06, DBLP:conf/rskt/Janicki08, journals/scw/Schulze11}.

For information retrieval tasks, Thomas and Hawking~\cite{DBLP:conf/cikm/ThomasH06} use pairwise comparisons in order to compare 
systems in real settings, where interactive retrieval is used in specific context
over ever-changing heterogeneous data. They show that click-through data highly correlates  with  
perceived
preference judgments. Sanderson et al.~\cite{DBLP:conf/sigir/SandersonPCK10} employ pairwise comparisons
with Amazon's Mechanical Turk~\cite{link:mturk} to obtain the correlation between user preference for text retrieval 
results and the effectiveness measures computed from a test collection.  The result of their study shows
that Normalized Discounted Cumulative Gain (NDCG) \cite{DBLP:journals/tois/JarvelinK02} is the measure that correlates best with
 the user perceived quality.
\begin{sloppy}
Using preference-based test collections is introduced by Rorvig~\cite{DBLP:journals/jasis/Rorvig90}
and later developed for text retrieval by Carterette et al.~\cite{DBLP:conf/ecir/CarteretteBCD08}. 
In this work, the authors show that preference judgments are faster to collect and provide higher levels of agreement, compared to
absolute relevance judgments. Preference-based effectiveness measures are proposed by Carterette and 
Bennett in \cite{DBLP:conf/sigir/CarteretteB08}, showing that they are stable and adhere to
the measurements based on absolute relevance judgments.
\end{sloppy}
Preference judgments between blocks of results are used by Arguello et al.~\cite{DBLP:conf/ecir/ArguelloDCC11} 
to evaluate aggregated search results. In this work, the small number of such blocks enabled the collection of preferences 
between  all pairs of blocks.  A suitable effectiveness measure in this case is the distance between the ranking
produced by the system and the reference ranking created based on the all-pair preferences. 
In our setting, the huge number of possible pairs prohibits an exhaustive evaluation, in which case the quality measure 
is more appropriate based directly on pairwise comparisons rather than using the reference ranking. 

For music similarity, Typke et al.~\cite{DBLP:journals/jdim/TypkeHNWV05} conclude that coarse levels 
of relevance measure, usually used in text retrieval, are {\it not applicable}. Instead,
they use a large number of relevance levels created from partially ordered 
lists.  The ground truth in this case is given as ranked list of document groups, such that documents in one group have 
the same relevance. The work by Urbano et. al.~\cite{DBLP:conf/ismir/UrbanoMML10} addresses some 
limitations of this approach by proposing different measures of similarity between groups of retrieved documents.
Measuring retrieval effectiveness with these large number of levels can be 
achieved using the Average Dynamic Recall~\cite{DBLP:conf/icmcs/TypkeVW06}.

Due to its low price and high scalability, crowd sourcing is a popular technique to obtain relevance 
assessments for information retrieval tasks~\cite{DBLP:conf/ecir/AlonsoB11, DBLP:conf/ecir/AlonsoST10, DBLP:conf/sigir/SandersonPCK10, DBLP:conf/sigir/KazaiMC09}.
The work by Alonso and Baeza-Yates~\cite{DBLP:conf/ecir/AlonsoB11} addresses the  design and implementation 
of assessments tasks in a crowd-sourcing setting, indicating that workers perform
as good as experts at TREC~\cite{link:trec} tasks. Similar results have also been achieved by Alonso et al.~\cite{DBLP:conf/ecir/AlonsoST10}
in the context of XML retrieval.  Snow et al.~\cite{DBLP:conf/emnlp/SnowOJN08}
show that Mechanical Turk workers were successful in annotating data for various natural language processing tasks, even
correcting the gold standard data for some specific tasks.

\section{Evaluation Dataset}
\label{sec:dataset}

A suitable evaluation dataset has to provide a wide coverage of both documents and queries. A common approach in 
traditional text retrieval is to use a large number of documents (e.g., obtained by crawling parts of the Web)
and to perform an initial filtering of documents based on existing approaches.

First,  existing approaches are used independently to retrieve the top ranked documents and then
these documents are combined (merged) to create, a so called,  ``pool'' of documents. 
Relevance assessments are then collected only for the documents in the pool in order to minimize the effort of the
human judges. This technique is commonly referred to as {\it pooling}. Due to the small number of existing soundtrack
recommendation approaches, pooling would result in a highly biased dataset. Hence, we have to assemble the set of queries (images) and 
documents (songs)  independently from the existing approaches in a way that ensures wide coverage while keeping the collection size tractable.\\

As defined above, an evaluation benchmark  $B=(Q, S, R)$ consists of a set of queries $Q$, a set of documents $S$ (that are songs in our case),
and a set of human relevance judgments $R$. The first step in creating the dataset is to select songs as documents and image sets 
as queries.

\subsection{Song Collection}

While building the song collection, we focus on popular music and try to achieve high coverage through  understanding 
common music aspects. There are two major aspects that people refer to when talking about music: the {\it feelings} induced by the music
and the {\it genre} it belongs to. We use Wikipedia~\cite{link:genreswiki} to obtain a  hierarchy of modern popular music genres and focus on
the genres that appear in the top level of the hierarchy. According to the creators of this Wikipedia page, music styles that are not commercially
marketed in substantial numbers are not included in the list.

Additionally, in order to avoid the complexity of working with a  large number of nation-specific music styles,
we eliminate genres specific to the origin of music, such as ``Brasilian music'' and ``Caribbean music''.  The resulting genres,
shown in Table~\ref{tb:genres}, range from Country and Blues, over Metal to Hip Hop and Rap.

\begin{table}[ht]
\centering
\begin{tabular}{@{}llll@{}}
\multicolumn{4}{l}{\bf Music Genres} \\
\midrule
Blues & Classical & Country & Easy listening \\
Electronic & Hip Hop and Rap & Jazz & Metal \\
Folk & Pop & Rock & Ska\\
\end{tabular}
\caption{List of music genres}
\label{tb:genres}
\end{table}

Next, we collect a set of feelings and organize them in two high-level groups: {\it positive and
negative feelings.} We obtained an exhaustive list of fine-grained feelings from {\it Psychpage}~\cite{link:psychpage}.
As the obtained list contains generic feelings, some are rarely conveyed by music, such as admiration, and satisfaction.
To identify feelings expressed through music we used the data from the {\it last.fm}~\cite{link:lastfm} music portal.
For each general feeling, we check how frequently an artist or a song is annotated with the tag (term) that describes a feeling, for instance, ``Sad''. 
This ``{\it wisdom of crowds}'' is gathered using Last.fm's search capabilities that retrieves all artists and songs annotated with a specific 
tag. While building the list, we employ a policy that a given feeling is not related to music if there are less than $500$ users who used 
this feeling as a tag. 
As the result, we get 7 positive and 7 negative feelings conveyed by music, shown in Table~\ref{tb:feelings}.
We see that  not only apparent feelings such ``Happy'' and ``Sad'' are there, but also
less frequent ones, such as ``Tragic'' and ``Optimistic'', are contained. This way, the number of feelings is limited while
still supporting high coverage.

\begin{table}[ht]
\centering
\begin{tabular}{@{}llll@{}}
\multicolumn{4}{l}{\bf Positive Feelings} \\
\midrule
Happy & Love & Calm & Peaceful \\
Energetic & Positive & Optimistic 
\vspace{5mm}
\end{tabular}
\begin{tabular}{@{}llll@{}}
\multicolumn{4}{l}{\bf Negative Feelings} \\
\midrule
Sad & Hate & Aggressive & Angry\\
Depressing & Pathetic & Tragic
\end{tabular}
\caption{Feelings induced by music}
\label{tb:feelings}
\end{table}

For each of the genres and feelings in the lists, we retrieve the top-10 played (listened to) artists. Again the {\it last.fm} portal is used for 
this task, as it contains the number of times an artist is listened to and enables the search for the top-K artists for a given query tag. 
For each artist, we acquire two representative songs, and automatically cut them to  30 seconds length---from minute 1:00  to 1:30.
As some artists appear in multiple groups, (e.g., in the ``easy listening'' genre and in ``optimistic'' feeling), 
the document collection consists of 470 songs in total. 
Having a total of 470 songs make a moderate collection size, while all major music genres and feelings are covered.

\subsection{Query Collection}

In the addressed soundtrack recommendation scenario, a query  is represented by a set of images. We create a list of $25$ queries, 
each containing 5 images, such that all images of a query follow a specific image theme. The initial list of image themes is retrieved 
from a list of photography forms, specified on Wikipedia~\cite{link:imagewiki}. For each of 
these themes, we retrieve images that are annotated with the theme, using the search functionality of 
Google's {\it Picasa}~\cite{link:picasa} photo sharing portal.
We manually inspect the returned results and use only themes that provided 
at least 5 coherent and meaningful images. This filtering step results in the final list
of $25$ image themes shown in Table~\ref{tb:imagethemes}. 
As we can see, image themes vary from photos taken underwater, over photos of people playing sports, to photos of special cloud forms. 
For each theme, a query is formed by manually selecting 5 publicly 
available images from Picasa, again keeping in mind the coherence and the meaningfulness of the image theme. 

\begin{table}[ht]

\centering
\begin{tabular}{@{}llll@{}}
\multicolumn{4}{l}{\bf Image Themes} \\
\midrule
Aviation & Architectural & Cloudscape & Conservation \\
Cosplay & Digiscoping & Fashion & Fine art\\
Fire & Food & Glamour & Landscape \\
Miksang & Nature & Old-time & Portrait \\
Sports & Still-life & Street & Underwater \\
Vernacular & Panorama & War & Wedding \\
Wildlife
\end{tabular}

\caption{List of query image themes}
\label{tb:imagethemes}
\end{table}

\section{Relevance Measure}
\label{sec:relevance}

Estimating the effectiveness of a retrieval engine is  
based on measuring the relevance of the returned results with respect to the given query.
In traditional text retrieval, relevance is represented by absolute judgments that usually make use of
a binary variable indicating that a document  is either relevant or not relevant to a given query. J{\"a}rvelin and 
Kek{\"a}l{\"a}inen~\cite{DBLP:conf/sigir/JarvelinK00} proposed a larger grading scale that allows for a finer separation
of relevant documents. We adopt such a fine-grained grading scale to assess the suitability of songs for series of images, extending
it to the extreme such that for each document (song) there is one level of relevance. Note that such fine-grained scales emphasize the point
of possible disagreement between human assessors, when determining how relevant a document is~\cite{DBLP:conf/sigir/Voorhees98}.

In the task of soundtrack recommendation, there is no such a notion of fulfilling a particular information need expressed by the query.
This renders the assessment less strict in the sense that in general  {\it all} songs can be used as background music.
That is, we do not explicitly have the notion of a document (song) being not relevant.
Further,  user perceived  relevance of a song with respect to images highly depends on knowledge of 
other available songs---it is a very relative assessment task: we can not simply present users small subsets
of songs and let them perform the assessment. A consistent full ranking of all available songs, for each query, is required.
Thus, we define the relevance $R(s|S,q)$ of the song $s$, given a song collection $S$, and a query $q \in Q$, as the rank of that 
document in the {\it perfect ranking}. With a ``perfect ranking'' we denote the full ranking that would be created by the 
``expert'' user. For a result list computed by a specific system for a given query, we can easily aggregate the relevance 
scores of the individual documents to obtain a final (non-zero) score.

A similar measurement is proposed for the task of similarity search in {\it sheet music}~\cite{DBLP:journals/jdim/TypkeHNWV05},
with expert users providing a full ranking of the documents. In contrast to our setup, there, it can indeed be decided if
two music sheets are completely not related (relevant to each other), which enables the use of pooling to obtain 
a filtered and shorter, list of documents for which the full ranking is done.

\subsection{Pairwise Preference Judgments}
\label{sec:pairwise_preference}

What remains is the problem of obtaining the full relevance ranking, for each benchmark query.
Doing this in an exhaustive way is  prohibitively expensive, though. Instead, the idea is to let users
evaluate a large number of song pairs, for each benchmark query.


We ask human judges  to evaluate a large number of song pairs, answering which one of the two presented songs 
fits better for a given query. This method of assessing is known as {\it preference judgments}. It is a convenient way to 
obtain relevance assessments, compared to obtaining absolute relevance judgments~\cite{DBLP:conf/ecir/CarteretteBCD08}. 
Ideally, the number of pairs judged for one query is large enough to reconstruct the whole ranking---which is
not achievable in practice. Thus, we collect judgments for only a subset of song pairs.

In addition to selecting the best out of two proposed songs, each judge is asked to assess how much better the selected song
fits to the query compared to the other song, on the scale from $1$ to $5$. A rating of  $1$ means  ``almost the same'' while 
$5$ means ``large difference''. The result of one human assessment is given in the following form $r=(q, s_{1}, s_{2}, p, d)$, 
where $q$ is the image theme query, $s_{1}$ and $s_{2}$ are songs, $p$ is the preferred song and $d$ is the difference 
between the songs. Optionally, assessors can provide a textual description (justification) of their decision.

The task at hand is, however, often  influenced by the  individual taste of the human judges---for  some queries more than for others. 
To capture this factor, we ask multiple assessors to judge the same song pair and use only the ones that show a high level of 
agreement. This way, the benchmark  can serve to evaluate generic soundtrack recommendation approaches.

To isolate the subjectivity of an individual assessor, based on the agreement level, we can check if the selection performed by 
the judges is {\it statistically significant}. In case the performed selection is statistically  significant we know that the agreement 
level between judges is high. In case selection is not statistically significant but there is still one song selected more then
the other, we can take this pair into consideration, keeping in mind that this was not an easy task---even for  human judges.

To check the statistical significance of the agreement between the judges, we formulate the following
null and alternative hypotheses:\\

$ \mathbf{H_{0}}${\bf : assessors are selecting songs randomly, i.e., do not consider the given query images}
\vspace{+2mm}

$ \mathbf{H_{1}}${\bf : assessors are selecting songs based on the given query images}\\

If the null hypothesis is true, each song (of a song pair) is independently selected with probability $p=0.5$. In that case,  the
songs are selected independently from the given query  and due to the independent trials we can calculate the probability of the 
final outcome using a binomial distribution. Applying a binomial test~\cite{statisticsPsy} gives us the probability of the outcome,
given that the null hypothesis is true. In case the probability of the assessment outcome is smaller than the required significance 
level (e.g., $\alpha=0.05$) we reject the null hypothesis and say that the agreement level for this question is statistically significant.

We create questions for human assessment by first creating song pairs in four different categories:
genre, positive, negative, and positive-negative. The pairs in the genre category are all song pairs of the
songs gathered based on the genre information. Similarly, the positive category contains all song pairs that have a positive 
feeling and negative category contains all song pairs with songs having a negative feeling. The positive-negative category 
consists of song pairs where one song is selected from the positive-feeling group and the second song is selected from 
the negative-feeling group. The first and the second song are shuffled before presenting them to the user 
to avoid an ordering bias.

Creating questions posed to human assessor is done by creating all possible triples where one element is 
an image-theme query and the other two are songs coming from  song pairs of one of the four categories created
in the previous step. All question triples are stored and the next question to be assessed by judges
is selected randomly among all non-assessed questions.

\subsection{System Effectiveness Measures}
\label{sec:effectiveness}

While collecting the assessments we had each question, i.e., song pair for a certain query, answered by 6 assessors.
Hence, the individual preference judgments need to be reconciled. To achieve this,
we compute the majority vote for each of the different agreement levels (four out of six (4/6), 5/6, and 6/6). Note that
for the agreement level 3/6 there is no majority vote, so we leave this level out.
For a certain agreement level, we  then obtain a set of relevance judgments $R$ with each $r\in R$ of the form $r=(q, s_{1}, s_{2}, p, d)$, where $q$ is 
an image query, $s_{1}$ and $s_{2}$ are songs, $p$ is the indication of the song preferred by the majority of users, and $d$
is the difference between songs averaged over multiple users assessing the same pair. 

Then, the quality (goodness) $G$ of a ranking can be computed using preference precision~\cite{DBLP:conf/sigir/CarteretteB08}
defined as:
\begin{equation}
G=\frac{ \mbox{\# correctly ordered pairs} }{\mbox{\# evaluated pairs}},
\end{equation}
where the pair of songs is correctly ordered if the song preferred by most assessors is located higher in the ranking compared 
to the other song, and the evaluated pairs are all song pairs that are assessed by the judges and are contained in the top-$K$ ranking.
A pair of songs is contained in the final top-$K$ ranking if at least one of the songs appears in the top-$K$ results. If only one song is in the top-$K$
results the rank of the second song is considered to be $K+1$. Intuitively, this measure rewards a system if its ranking agrees with a user's perceived 
preference, resulting in a higher value with a higher agreement between the two.

Although $G$ is normalized to the  $[0, 1]$ interval, due to  possible inconsistencies in the transitivity caused by pairwise comparisons, 
this interval might shrink. An example of the inconsistency can be seen in three pairwise comparisons between the three elements, $\{a,b,c\}$,
where $a$ is preferred to $b$, $b$ preferred to $c$, but $c$ is preferred to $a$. We see that it is impossible to create a ranking satisfying 
all comparisons so the upper bound is lower than $1$, and the lower bound is higher than $0$. Of course, this kind of situations arise as
pairwise comparisons are created independently from each other and potentially by other assessors. Still, the actual lower and upper 
bounds can easily be computed once all preference judgments are collected. 

\subsubsection*{Weighted Effectiveness Measure}
\label{sec:difference}

The specified  difference between the songs, denoted as  $d$, can be considered as the strength of preference  and, hence, can be taken 
into account when assessing the quality of a system. As multiple judges are evaluating the same pair of songs for a given query, the final 
value of the difference  between songs is taken as the average of the single evaluations.\\

The obvious way to extend the preference precision measure using the preference strength is as follows:
\begin{equation}
G_w=\frac{ \sum_{\mbox{correctly ordered pairs}}p_s}{\sum_{\mbox{evaluated pairs}}p_s}
\end{equation}
where $p_s$ is the preference strength, having higher value if the preference is stronger. For instance, the preference is strong toward 
one song if the difference between the two songs is large. Thus, we can use this difference between songs directly as preference strength.
Clearly, this measure gives more weight to the preference judgments which were obvious for humans, and dampens the effect of judgments 
for which even the assessors were not sure about their preference.

\section{The Benchmark}
\label{sec:building}

Processing large amounts of human-involved tasks can be efficiently addressed
using Amazon's Mechanical Turk~\cite{link:mturk}. This service represents a mediator between
the requester---a person or an organization posting tasks to be done---and a number of workers---people willing to perform
these tasks, while getting paid for it. 

There are studies~\cite{DBLP:conf/ecir/AlonsoB11, DBLP:conf/ecir/AlonsoST10, DBLP:conf/sigir/SandersonPCK10} on 
the usage of the Mechanical Turk service to collect relevance assessments. All of these studies face the same problem: 
determining whether the worker really prefers a selected document (song), or if the selection
is done randomly to simply gain money, without spending sufficient effort on the assessment task.

To remove  assessments of such ``cheaters'', a certain set of question with known answers is inserted
in the evaluation task. These questions are referred to as ``trap questions'', ``honey pots'' or ``gold standard'' questions. 
Creating trap questions for text retrieval, in case of preference judgments, is an easy task: a pair of one relevant and one 
obviously irrelevant document are presented to the evaluator. Cheating evaluator are then identified by the percentage of 
times the obviously irrelevant document is selected as the preferred answer.

As our task is more prone to subjectivity, we collect judgments in two phases. 
In the first phase, we build a set of ``gold standard'' questions by collecting 
judgments from students on our campus, in the controlled environment of our offices.
These ``gold standard'' questions are used as trap questions for the second phase of assessments gathering, using Mechanical Turk workers.
The main hypothesis behind this approach is that evaluators (workers) employed in our offices would have less
incentives to cheat as we pay them by hour, not by the number of performed assessments.

Each question (song pair and query image theme) is answered by six assessors. We chose six assessors, 
as the significance level of $\alpha=0.05$ is achieved in case when all six assessors agree on the preferred song.
Only the questions with an agreement level of ``six out of six'' are used to create trap questions 
for the next phase---considering only the preferred song, not the level of difference $d$.
The probability of achieving this level of agreement randomly is quiet low, with a p-value of $0.03125$, 
which makes it safe to use these questions  as trap questions. 

Note that at the evaluation phase, we can choose to perform the quality assessment using only the second phase 
assessments, obtained from Mechanical Turk, to indisputably avoid a potential student population bias. 

\subsubsection*{Collecting through Mechanical Turk}
Collecting a larger amount of assessments is achieved in the second phase, with assessments being made by Mechanical 
Turk workers. To obtain a robust benchmark, again, each question is answered by six workers. This enables a later
evaluation based on different levels of agreement.

{\it Cheating} workers are identified as the ones that have performed a large number of questions---expecting high
money reward---while choosing answers at random. Due to the binary nature of the questions, cheaters answer 
approximately only 50\% of all trap questions correctly. We used a threshold of at least $100$ answered questions and less 
than 65\% correct trap questions to {\it reject} a work of a {\it cheating} worker. We used one trap question per five 
regular questions.

Because workers prefer small tasks~\cite{DBLP:conf/ecir/AlonsoB11}, we created one HIT (Human Intelligent Task) for each question.
We set the reward to $\$0.02$ for each performed task, as the reward per task has only a small impact on the quality but rather influence 
the quantity of the performed tasks~\cite{DBLP:conf/kdd/MasonW09}.

\subsection{Benchmark Statistics}
\label{sec:statistics}

To obtain trap questions for the Mechanical Turk workers,
we collected assessments from $30$ students. Students were able to 
choose whether they want to participate in the study for one hour or two hours, while being payed on an hourly basis.
$29$ out of $30$ students participated in the study for two hours, resulting in the 
total of $665$ questions, each question assessed by 6 students. 

\begin{figure}[t]
\centering
\includegraphics[width=0.9\columnwidth]{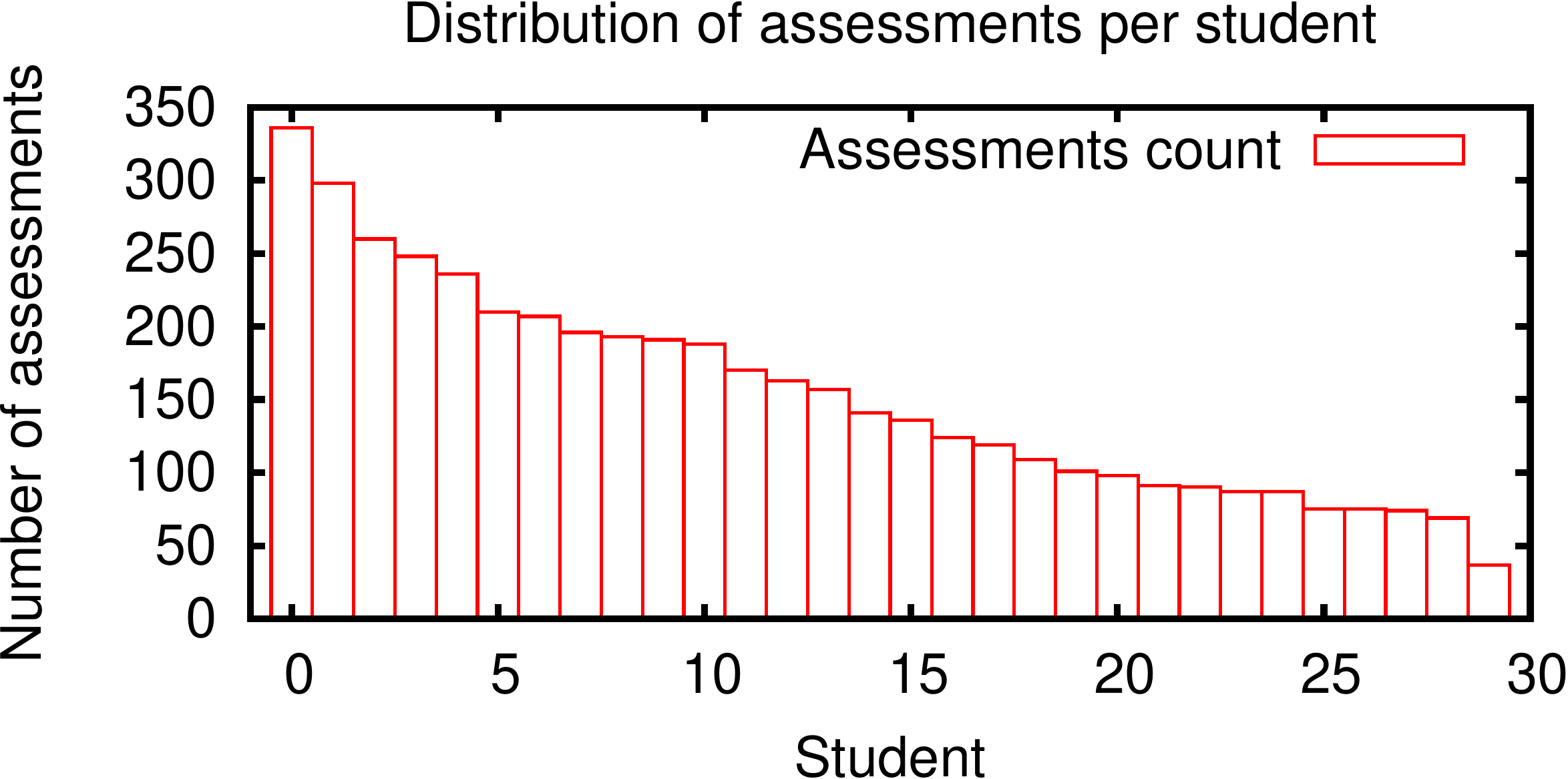}
\caption{Distribution of assessments per student}
\label{fig:evalsStudents}
\end{figure}

Figure~\ref{fig:evalsStudents} shows the number of assessments per student, sorted in descending order.
We observe a high variance in assessment performance, even if we exclude the student that assessed songs
for only one hour. Using Pearson's correlation coefficient, we investigate if this variance comes 
from the open ended question, used to elaborate their decision. Pearson's correlation coefficient between total text length (word count) and 
a number of assessments is only $-0.0875$, indicating that typing the explanation answer is not the reason for 
the variance in individual assessment performance. If we characterize the agreement with other assessors 
as a quality estimate of the assessor, we can also check if assessors that produced a large number of assessments 
have a drop in quality, i.e., have low agreement with others.
Pearson's correlation coefficient between assessments made and agreement with other assessors is only
$0.04141$ indicating that the quality of the work is also independent of the assessors performance.

\begin{table}[ht]
\centering
\begin{tabular}{@{}lrrrr@{}}
\toprule
& \multicolumn{4}{c}{\bf Agreement level} \\
& 3/6 & 4/6 & 5/6 & 6/6 \\
\midrule
percentage & 12.33 & 32.18 & 26.17 & 29.32 \\
average difference & 2.75 & 2.90 & 3.11 & 3.65 \\
\bottomrule
\end{tabular}

\caption{Agreement levels for student assessors}
\label{tb:agreementLocal}
\end{table}

Table~\ref{tb:agreementLocal} shows the percentages of questions with different agreement levels for student
evaluations. Agreement level ``x/y'' means that x out of y assessors agreed on one song.
The observed values in Table~\ref{tb:agreementLocal} suggest that we can reject the null hypothesis 
that student answers were done by randomly choosing songs, supported by the Chi-square test $\chi^{2}=1586.86$, $df=3$, $p<0.0001$.
This level of significance shows that such a high level of agreement between assessors is almost impossible to 
achieve by pure chance, but that the task at hand is reasonable and meaningful for the assessors.
As we can see, 29.32\% (i.e., 195) of all questions  have agreement level of 6/6, making them 
applicable as trap questions for the second phase. The global agreement
statistics shows that student assessors agree with each other in 66.94\% of all cases. This is lower than 
the agreement level for the traditional text retrieval task (75.85\%)~\cite{DBLP:conf/ecir/CarteretteBCD08}, 
which indicates that the task of soundtrack recommendation is more subjective.

The averaged difference between songs is also reported for student evaluations in Table~\ref{tb:agreementLocal}.
We see that the average difference is smallest, $2.75$, for the
questions with low agreement level and gradually increases to $3.65$ for the questions with the {\it six
out of six} agreement level. Pearson's correlation coefficient between agreement level and the average
difference between songs is $0.3556$, with a randomization test ($100.000$ permutations) showing that 
this result is not achievable randomly, $p < 0.0001$. 

\begin{figure}[t]
\centering
\includegraphics[angle=-90, width=0.95\columnwidth]{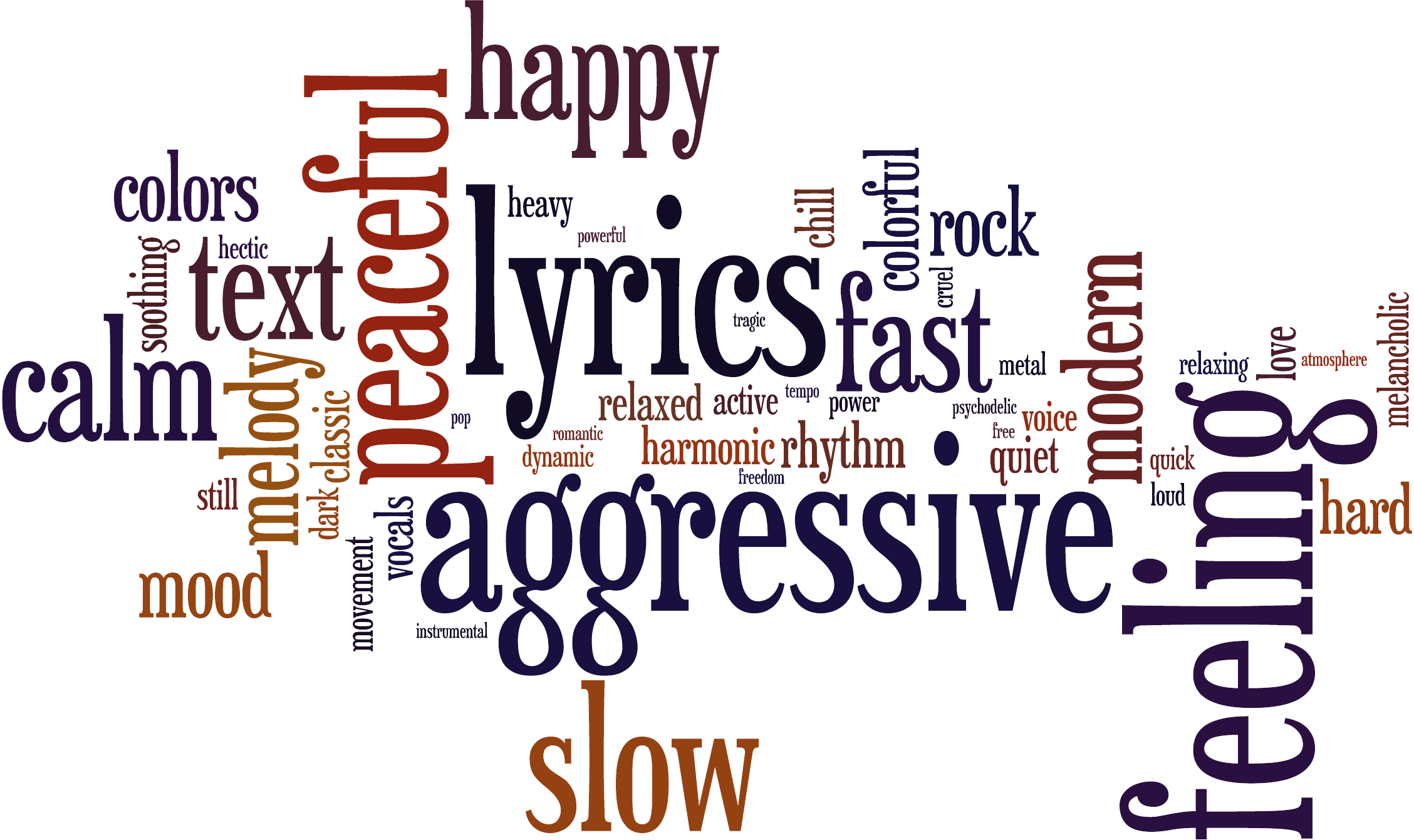}
\caption{Concepts used to describe matching between images and music}
\label{fig:concepts}
\end{figure}

While assessing the songs, assessors were able to provide textual description of their decision. We analyzed
the collected descriptions to see which concepts assessors use for music and images when they are
being matched together. We manually extracted all concepts from the descriptions collected from students, 
shown in Figure~\ref{fig:concepts} where the size of a word represent its frequency in the text.\\

In the second phase, we used Amazon Mechanical Turk to collect a larger number of assessments. 
Our aim was to collect enough assessments such that each song for each image query had a chance
of being judged once. This required us to have more than $5875$ questions evaluated, each question
assessed by six assessors. In the end, we collected assessments evaluating $5990$ questions in total. 

Overall, we had  $269$ assessors participating in the study. On average, each of them performed $138.69$ assessments. 
As there was no time limit for each assessor, the skew in the number of performed assessments is much larger
than for the students in phase one, ranging from one evaluation up to $3845$ evaluations per assessor. Gold standard
questions enabled us to detect $15$ {\it cheating} workers and to reject their work, being
replaced by other workers' assessments.

\begin{table}[ht]

\centering
\begin{tabular}{@{}lrrrr@{}}
\toprule
& \multicolumn{4}{c}{\bf Agreement level} \\
& 3/6 & 4/6 & 5/6 & 6/6 \\
\midrule
percentage & 17.15 & 33.89 & 28.60 & 20.37 \\
average difference & 3.09 & 3.17 & 3.37 & 3.68 \\
\bottomrule
\end{tabular}

\caption{Agreement levels for Mechanical Turk workers}
\label{tb:agreementMturk}
\end{table}

The percentage of questions with respect to agreement levels for Mechanical Turk workers is shown in Table~\ref{tb:agreementMturk}.
As we can see, the percentages of questions with high-agreement levels are lower than for student assessments.
Still, we can safely reject the hypothesis of randomly provided answers, with $\chi^{2}= 6605.18$, $df=3$, $p<0.0001$.
The reduction in the agreement level might also be an effect of the more diverse population of workers compared to the 
population of students.

We see that the percentage of questions with agreement level of ``5/6'' and ``6/6'' is close to 50\%, which renders
almost half of the evaluated questions usable with high confidence. Overall,  the agreement between
mechanical turk workers was achieved in 62.10\% of all cases, slightly less than the overall agreement of students,
which corresponds to the drop in the number of high-agreement questions.

Again we see that there is a correlation between average difference between the songs and the agreement level. Pearson's
correlation coefficient in this case is $0.2928$, with randomization test ($100.000$ permutations) 
showing again that the probability of randomly achieving this value is $p < 0.0001$. 

\subsubsection{Query Type Statistics}
In this section, we report on the statistics of the assessments concerning question types and image themes. 

After merging evaluations performed by students and by the Mechanical Turk workers we calculated the
percentage of questions at different levels of agreement for each of the question types, shown in 
Table~\ref{tb:agreementSubgroupsMturk}. The average difference between songs is also reported for each
agreement level and each question type.

As we can see, the largest percentage of high-agreement questions is achieved for questions where both 
songs have a negative feeling. Inspecting the assessments for these questions revealed that melancholic songs
with slow rhythm were usually preferred to fast, loud, and aggressive songs. It is interesting to see that
questions formed from different music genres had the least amount of high agreement. This might indicate
that songs from different genres might not always be largely different, or that assessments were biased 
towards preferred music genre, which could be a cause of disagreements. 

\begin{table}[ht]
\centering
\begin{tabular}{@{}llrrrr@{}}
\toprule
& & \multicolumn{4}{c}{\bf Agreement level} \\
& & 3/6 & 4/6 & 5/6 & 6/6 \\
\midrule
\multirow{2}{*}{Genres} & percentage & 18.24 & 37.51 & 28.30 & 15.95 \\
& avg. difference & 3.16 & 3.23 & 3.37 & 3.65 \\
\cmidrule(l){2-6}
\multirow{2}{*}{Positive} & percentage & 17.28 & 35.57 & 29.27 & 17.88 \\
& avg. difference & 2.98 & 3.06 & 3.24 & 3.56 \\
\cmidrule(l){2-6}
\multirow{2}{*}{Negative} & percentage & 13.89 & 30.01 & 28.84 & 27.26 \\
& avg. difference & 3.00 & 3.10 & 3.37 & 3.67 \\
\cmidrule(l){2-6}
\multirow{2}{*}{Pos.-Neg.} & percentage & 17.35 & 31.85 & 26.95 & 23.85 \\
& avg. difference & 3.10 & 3.17 & 3.43 & 3.79 \\
\bottomrule
\end{tabular}

\caption{Statistics by question type}
\label{tb:agreementSubgroupsMturk}
\end{table}

The percentage of questions with {\it five out of six} and {\it six out of six} agreement levels together with the
average difference between songs is shown for each image theme in Table~\ref{tb:imageTheme}:
The average difference between songs does not change a lot over different image themes,
varying from $3.19$ for architecture up to $3.41$ for fashion and wedding themes. On the other
hand, the number of high-agreement questions varies substantially, ranging from $35.7$\% for the war 
theme to $60.9$\% for the fine art theme. As expected, emotionally intense themes such as the war, fire, and aviation
themes have a substantially lower level of agreement than the ``calm'' themes such as fine art, portrait, and nature. 

\begin{table}[ht]
\centering
\begin{minipage}{0.49\columnwidth}
\begin{tabular}{@{}lrr@{}}
\toprule
Image theme & Agr.& Diff.\\
\midrule
architecture & 41.2 & 3.19\\
aviation & 41.4 & 3.26\\
cloudscape & 49.8 & 3.23\\
conservation & 44.9 & 3.28\\
cosplay & 42.1 & 3.26\\
digiscoping & 55.5 & 3.34\\
fashion & 45.0 & 3.41\\
fineart & 60.9 & 3.29\\
fire & 38.1 & 3.30\\
food & 53.9 & 3.33\\
glamour & 47.4 & 3.27\\
landscape & 51.7 & 3.24\\
miksang & 51.8 & 3.32\\
\bottomrule
\end{tabular}
\end{minipage}
\begin{minipage}{0.49\columnwidth}
\begin{tabular}{@{}lrr@{}}
\toprule
Image theme & Agr.& Diff.\\
\midrule
nature & 58.8 & 3.34\\
old-time & 54.1 & 3.27\\
panoram & 51.6 & 3.23\\
portrait & 60.1 & 3.40\\
sports & 42.1 & 3.29\\
still life & 48.2 & 3.26\\
street & 40.0 & 3.29\\
underwater & 58.8 & 3.39\\
vernacular & 52.1 & 3.26\\
war & 35.7 & 3.36\\
wedding & 58.6 & 3.41\\
wildlife & 53.6 & 3.30\\
\bottomrule
\end{tabular}

\end{minipage}
\caption{Statistics by image theme}
\label{tb:imageTheme}

\end{table}

\section{Evaluating State-of-the-art}
\label{sec:evaluation}

To go beyond the plain proposal of a benchmark, we now present its application to the evaluation of the two state-of-the-art approaches in the area of
soundtrack recommendation. First, an approach by Li and Shan~\cite{DBLP:conf/mm/LiS07} that
is based on emotion detection in images and music---we refer to this approach as the {\it emotion-based approach}. 
Second, our approach, coined Picasso~\cite{DBLP:conf/sigir/StuparM11}, that 
extracts information from publicly available movies and uses that information to create a match between images 
and music.

\subsection{Approaches}
\label{sec:approaches}

{\bf Emotion-based Approach:} The approach by Li and Shan~\cite{DBLP:conf/mm/LiS07}, was originally developed to recommend music for impressionism 
paintings, but can in general be applied to arbitrary sets of images. The key idea is to detect emotions in both images and 
music and to employ this information for the match making. The detection of emotions and the recommendation of music is 
done through methods based on a graph representation of multimedia objects, called the {\it mixed media graph}.

In the mixed media graph, each multimedia object and the associated attributes are represented as vertices, where attribute vertices
are either labels associated with the object or the low-level features extracted from the object. Object vertices are connected 
to the corresponding attribute vertices, with additional edges created between the vertices containing low-level features,
based on K-Nearest Neighbor search~\cite{MultiDimensionalStructures}: for each feature vector, edges are created to 
its $N$ closest neighbor vertex.

\begin{figure}[t]
\centering
\includegraphics[width=0.95\columnwidth]{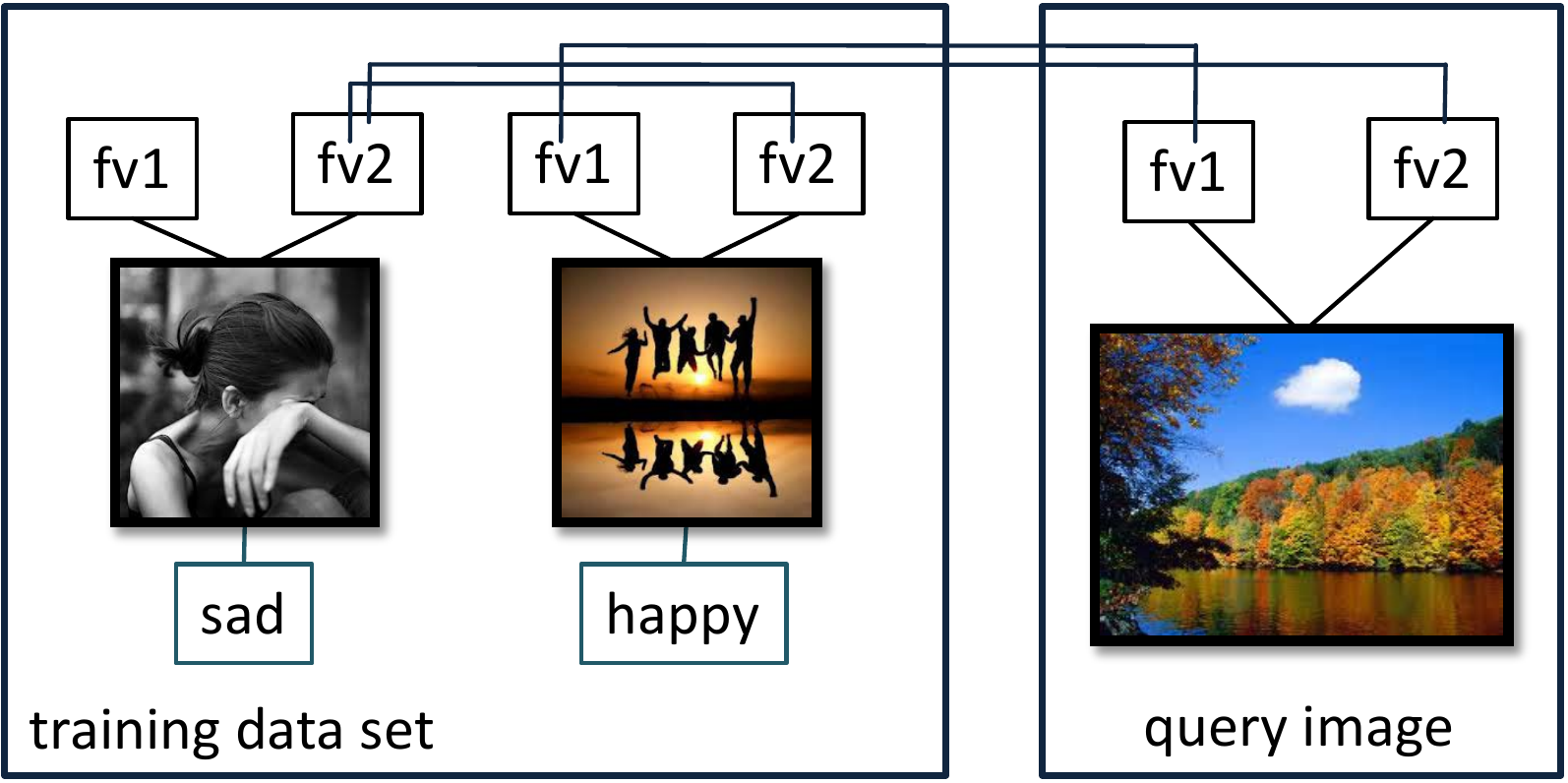}
\caption{Emotion-based approach: detecting emotions in query images}
\label{fig:emotionbased}
\end{figure}

To detect emotions in given query images, a training set of images with labeled emotions is represented in the mixed 
media graph, as illustrated in Figure~\ref{fig:emotionbased}. After introducing nearest-neighbor edges, a {\it random walk 
with restarts} is applied to find the labels, i.e., emotions, with the largest weight.

The mixed media graph is again used in the second step of the soundtrack recommendation---this time with songs as multimedia 
objects. In this step a ``dummy'' object is created as a query, with emotions from the previous step as labels. Once the edges 
are created between the labels, again the random walk with restarts is applied but this time with the aim of finding the songs 
with the highest weight. The songs from the collection are recommended in decreasing order of their weight.\\

{\bf Picasso:} The recommendation process in Picasso~\cite{DBLP:conf/sigir/StuparM11} is based on information 
extracted from publicly available movies. This extraction is done in a preprocessing phase through the following eight steps:
\begin{itemize}
  \setlength{\itemsep}{2pt}
  \setlength{\parskip}{0pt}
  \setlength{\parsep}{0pt}
\item[(i)] {the soundtrack of the movie is extracted}
\item[(ii)]{music/speech classification is done on the soundtrack}
\item[(ii)]{speech parts are discarded}
\item[(iv)]{screenshots, during the music parts, are taken}
\item[(v)]{parts of the same scene are detected}
\item[(vi)]{the soundtrack is split according to the scenes}
\item[(vii)]{for each soundtrack part the distance to all songs is calculated}
\item[(viii)]{the song lists are sorted in increasing order of distance}
\end{itemize}

The result of the extraction is an index that contains $<$movie screenshot, soundtrack part$>$-pairs, where the soundtrack
part is the one that is surrounding that screenshot. For each of these pairs, the index additionally contains a sorted list of songs by
their similarity to the soundtrack part. 

When an image is submitted to the system, Picasso finds the $K$ most similar movie screenshots to the given query image.
It then retrieves the $K$ most similar songs to the soundtrack parts that corresponds to the retrieved screenshots. After the song 
lists are retrieved, smoothing is applied to dampen the effect of outliers and the final score for the song is calculated.

Having multiple images submitted to the system, the recommendation process starts by recommending songs individually for each 
image. The problem is then to combine the lists of songs for each individual recommendation into a single recommendation list.
Picasso casts this problem into a group recommendation problem~\cite{Amer-Yahia:2009:GRS:1687627.1687713} and uses the 
established approaches to solve it. The casting is achieved by representing the images as users, and song lists as their preferences.

\subsection{Evaluation Results}
\label{sec:results}

For the emotion-based approach to operate we need two training datasets, that is, a set of images and a set
of songs with labeled emotions. As part of the songs in the benchmark were acquired based on their emotion labels,
we already have a training dataset for the songs. 

As a training dataset for images we use the International Affective Picture System~\cite{LangBradleyCuthbert99} 
dataset. It contains $1196$ images, each placed on the three dimensional space of emotions it evokes.
The three dimensional space consists of two primary dimensions, namely valence---ranging from pleasant to unpleasant---and 
arousal---ranging from calm to excited. Third less strongly related dimension represents a dominance expressed in 
the image. For more details on this emotion space representation see~\cite{russell1980circumplex}.

\begin{figure}[ht]
\centering
\includegraphics[width=0.75\columnwidth]{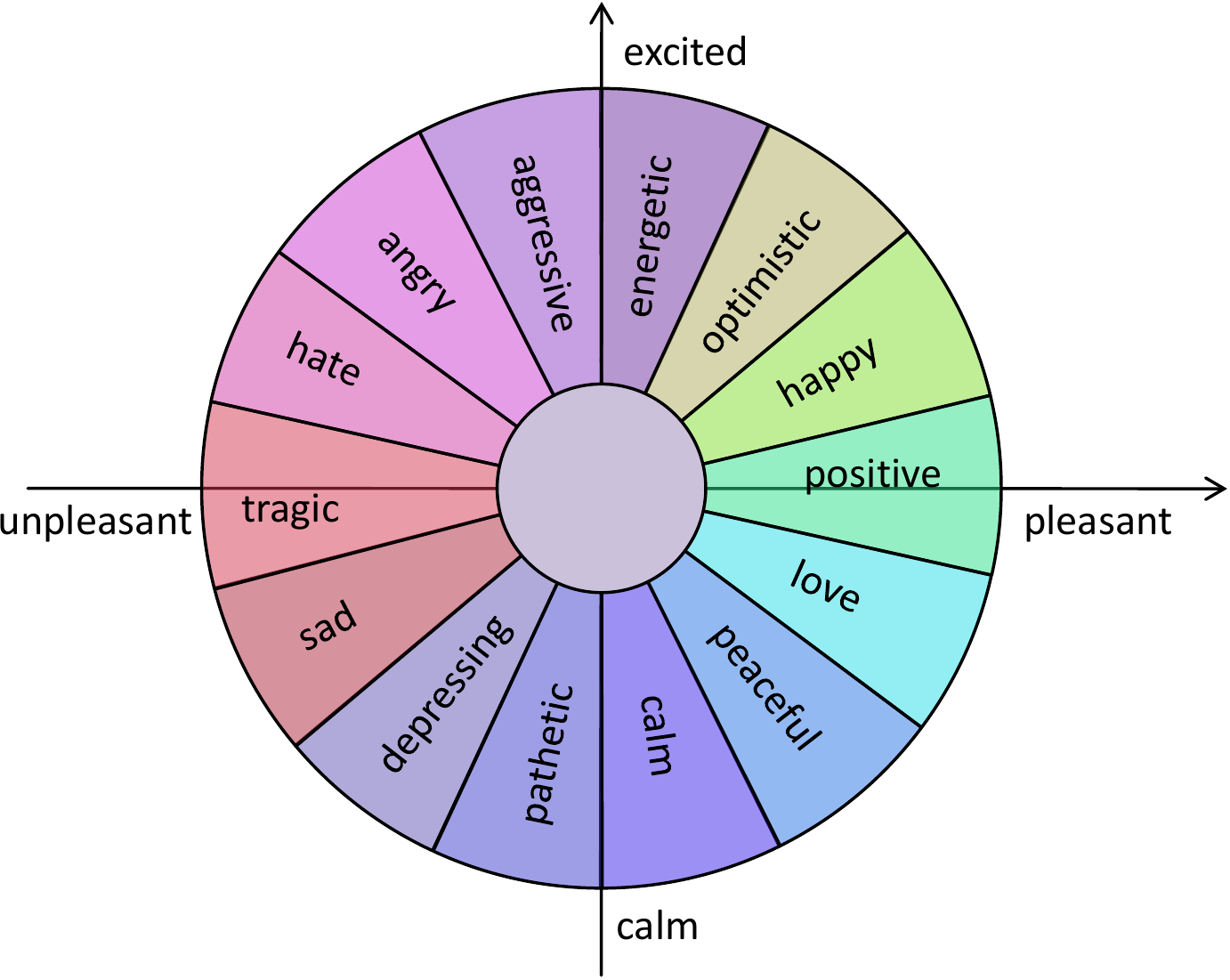}
\caption{Mapping emotion labels to two dimensional emotion space}
\label{fig:mapping}
\end{figure}

To create a match between music and images, we need a unified representation of emotions. This is achieved
by mapping emotions, used to label music, into the three dimensional space of emotions, used to label images.
Each image is labeled by one emotion, where the emotion label corresponds to the area of the space indicated by the 
two primary dimensions valence and arousal. The used mapping is shown in Figure~\ref{fig:mapping}.

To create the index for Picasso, we extracted information from 50 publicly available movies.
All the movies  originate from Hollywood production but cover a wide variety in genres and styles. 
In total, the final index contains $10,454$ snapshots taken and the same number of corresponding 
soundtrack parts. 

We execute both systems for each of the $25$ queries from the benchmark requesting the top-20 songs 
as a recommendation result. 

\begin{table}[ht]

\centering
\begin{tabular}{@{}lrrr@{}}
\toprule
& \multicolumn{3}{c}{\bf Agreement levels} \\
\cmidrule(l){2-4}
System & 6/6 & +5/6 & +4/6 \\
\midrule
Emotion-based & 0.658 & 0.595 & 0.559 \\
Picasso & 0.782 & 0.690 & 0.614 \\
\hline
\hline
Fisher's exact test (two-tailed) & 0.0530 &  0.0249 & 0.0938 \\
\bottomrule
\end{tabular}

\caption{Preference precision}
\label{tab:precision}
\end{table}

The preference precision for both systems is shown in Table~\ref{tab:precision}. The first column contains
the preference precision measures when the systems are evaluated using only the questions with {\it six out of six}
level of agreement. Further, adding questions with {\it five out of six} agreement level to the evaluation results
in precision shown in the second column, and finally, the evaluation with {\it four out of six} agreement level questions
added is shown in the third column. 

Fisher's exact test is used to examine the probability of achieving these differences in precisions in case the results
come from the {\it same} system (hypothetically). The contingency tables for the Fisher's exact test are created by 
counting the number of correctly and incorrectly ordered pairs for both approaches.

We see that both systems perform best when the questions used for evaluation are the ones
for which assessors agreed on the answers. The performance of both systems drops when 
questions, for which users did not easily agree on the answers, are added to the evaluation. The achieved precision numbers 
indicate that Picasso performs better with regard to questions at all levels of agreement. Fisher's exact test shows that 
it is not likely that this difference in precision is achieved by chance. Although the systems achieve precision
up to $0.782$ (Picasso system for {\it six out of six} agreement level) there is still a large space for improvements in 
both systems.

We calculate also the weighted preference precision that takes into account the difference between songs specified
by the assessors. As the difference between songs is bigger when one song fits a lot better to the query we put more 
emphasis on these song pairs to reward/penalize a system for correct/incorrect ordering of these pairs. 

\begin{table}[ht]

\centering
\begin{tabular}{@{}lrrr@{}}
\toprule
& \multicolumn{3}{c}{\bf Agreement levels} \\
\cmidrule(l){2-4}
System & 6/6 & +5/6 & +4/6 \\
\midrule
Emotion-based & 0.667 & 0.607 & 0.570 \\
Picasso & 0.818 & 0.728 & 0.645 \\
\hline
\hline
Student's t-test (two-tailed) & 0.0148 &  0.0042 & 0.0197\\
\bottomrule
\end{tabular}

\caption{Weighted preference precision}
\label{tab:weighted}
\end{table}

The weighted preference precision of both systems is shown in Table~\ref{tab:weighted}. As we can see, the weighted precision 
for both systems is higher than the preference precision. This shows that incorrectly ordered song pairs were the ones with a small 
difference between the songs. Again, the best precision is achieved for high agreeing questions as the number of correctly ordered 
song pairs is higher. We also  see that Picasso performs better than the emotion-based approach. By calculating student's t-test, also 
shown in Table~\ref{tab:weighted}, with positive differences for correctly ordered pairs and negative for incorrectly ordered
ones, we can reject the hypothesis the means for the two systems are the same. 

\section{Conclusion}
\label{sec:conclusion}

In this work, we  addressed the problem of building a comprehensive and reusable benchmark
for soundtrack recommendation systems. We formally defined the task of soundtrack recommendation
and the format of the evaluation benchmark. Assessments were collected in form of preferences 
judgments: In the first phase from the students at the university and in the second phase through 
Amazon's Mechanical Turk. We presented detailed statistics for collected assessments with respect to the 
agreement levels between assessors and different query types. We showed how the obtained judgments 
can be used to evaluate the quality of the soundtrack recommendation engines and reported on the 
performance of the state-of-the-art approaches. 

\balance

\bibliographystyle{IEEEtran}

\end{document}